\documentclass{CUP-JNL-EDS}

\usepackage{latexsym}
\usepackage{graphicx}
\usepackage{multicol,multirow}
\usepackage{amsmath,amssymb,amsfonts}
\usepackage{mathrsfs}
\usepackage{amsthm}
\usepackage{rotating}
\usepackage{appendix}
\usepackage[authoryear]{natbib}
\usepackage{ifpdf}
\usepackage[T1]{fontenc}
\usepackage{times}
\usepackage{newtxmath}
\usepackage{textcomp}%
\usepackage{xcolor}%
\usepackage{hyperref}
\usepackage{lipsum}
\usepackage{xcolor}

\articletype{APPLICATION PAPER}
\jname{Environmental Data Science}
\jyear{2026}
\jvol{4}
\jdoi{10.1017/eds.2020.xx}

\begin{document}

\begin{Frontmatter}

\title[Article Title]{IPSL-AID: Generative Diffusion Models for Climate Downscaling from Global to Regional Scales}

\author[1]{Kishanthan Kingston}
\author[1]{Olivier Boucher}
\author[2]{Freddy Bouchet}
\author[1,2]{Pierre Chapel}
\author[1]{Rosemary Eade}
\author[4]{Jean-Francois Lamarque}
\author[3]{Redouane Lguensat}
\author*[1]{Kazem Ardaneh}\email{kardaneh@ipsl.fr}\orcid{0000-0003-0473-6907}
\address[1]{\orgdiv{Climate Modeling Center}, \orgname{Sorbonne University, CNRS, IPSL}, \orgaddress{\city{Paris}, \country{France}}}
\address*[2]{\orgdiv{LMD}, \orgname{CNRS, ENS/PSL, Sorbonne University, IPSL}, \orgaddress{\city{Paris}, \country{France}}}
\address*[3]{\orgdiv{Climate Modeling Center}, \orgname{Sorbonne University, IRD, IPSL}, \orgaddress{\city{Paris}, \country{France}}}
\address*[4]{\orgdiv{Climate Modeling and Analysis}, \orgname{LLC}, \orgaddress{\city{Superior}, \state{CO},  \country{USA}}}
\received{---}
\revised{---}
\accepted{---}

\authormark{Kingston et al.}

\keywords{Climate downscaling, Generative models, Diffusion models, Uncertainty quantification, Machine learning}

\abstract{
Effective adaptation and mitigation strategies for climate change require high-resolution projections to inform strategic decision-making. Conventional global climate models, which typically operate at resolutions of 150 to 200 kilometers, lack the capacity to represent essential regional processes. IPSL-AID is a global to regional downscaling tool based on a denoising diffusion probabilistic model designed to address this limitation. Trained on ERA5 reanalysis data, it generates 0.25$^{\circ}$ resolution fields for temperature, wind, and precipitation using coarse inputs and their spatiotemporal context. It also models probability distributions of fine-scale features to produce plausible scenarios for uncertainty quantification. The model accurately reconstructs statistical distributions, including extreme events, power spectra, and spatial structures. This work highlights the potential of generative diffusion models for efficient climate downscaling with uncertainty estimation.}

\policy{{ IPSL-AID is a generative tool designed for climate downscaling—from the global to the regional scale—developed through a collaboration between computer scientists, AI/ML specialists, and climate scientists. This interdisciplinary team offers a multidimensional perspective that balances computational efficiency, statistical rigor, and physical consistency, hence making the work accessible to researchers across these fields.
}}

\end{Frontmatter}

\section{Introduction}

Anthropogenic climate change poses substantial risks to critical socio-economic sectors, such as agriculture, forestry, energy, and water supply. The development of effective adaptation and mitigation strategies requires accessible, high-resolution climate projections to anticipate impacts at both local and regional scales. General circulation models (GCMs), which are widely used in climate research, typically operate at spatial resolutions of approximately 150–200 km \citep{ipcc_2023}. This coarse resolution is insufficient to capture essential fine-scale processes, especially those shaped by topography, land-sea contrast, and surface heterogeneity, all of which are vital to regional climate dynamics and extreme weather events. Therefore, downscaling climate model outputs to finer resolutions is necessary to provide relevant climate information at the local level.

Weather and climate downscaling methods are typically classified into two categories \citep{hewitson_1996, wilby_1997}. (i) Dynamic downscaling uses nested regional climate models (RCMs) to explicitly simulate physical processes at finer spatial scales \citep{giorgi_1991, vrac_2012, giorgi_2015}. However, the high computational demands of this approach limit its feasibility for large ensembles or long simulation periods. (ii) Statistical downscaling establishes empirical relationships between large-scale climate predictors, such as pressure fields and circulation indices, and local-scale variables \citep{wilby_2004, vrac_2012, maraun_2018}. However, this approach has limited capacity to capture complex nonlinear interactions and to maintain multivariate physical consistency within the generated fields.

Recent studies have shown that RCMs can be emulated with relatively low computational cost \citep{doury_2023, rampal_2024}. Artificial intelligence-based methods for weather prediction, including Pangu-Weather \citep{bi_2023}, GraphCast \citep{lam_2023}, and FourCastNet \citep{pathak_2022}, have demonstrated significant potential \citep{koldunov_2024}. These data-driven models facilitate efficient, automated downscaling by learning to incorporate fine-scale details from reanalysis datasets such as ERA5. However, these approaches are generally deterministic and may introduce systematic biases. Critically, they often lack a robust framework for quantifying the uncertainty inherent in the downscaling process, which is essential for comprehensive risk assessment. {Recent works in downscaling have highlighted the potential of diffusion models \citep{rampal_2024}. CorrDiff, introduced by \citet{mardani_2025}, is a residual corrective diffusion model that downscales 25~km ERA5 inputs to 2~km resolution over Taiwan, demonstrating strong performance in reconstructing spatial spectra and synthesizing radar reflectivity. Similarly, \citet{watt_2024} used a diffusion model to downscale 2$^{\circ}$ ERA5 fields to 0.25$^{\circ}$ over the continental United States, exhibiting superior performance compared to U-Net and linear interpolation. While these studies confirm the promise, they are constrained to fixed regional domains. Moreover, neither study addressed precipitation downscaling, a critical variable for impact assessment, and both relied on region-specific training that cannot be generalized to new geographic areas without retraining.}

{This paper introduces IPSL AI Downscaling (IPSL-AID), a tool for global to regional downscaling using denoising diffusion probabilistic models \citep{ho_2020_ddpm, song_2020, song_2020_ddim, karras_2022} with several key innovations. First, we propose a \textit{random block sampling strategy} that enables training the model on global data with limited GPU resources, thereby delivering a model applicable to arbitrary regions without retraining, representing a step toward foundation models for downscaling. Second, the model architecture processes \textit{multiple time steps simultaneously} (with a batch size of 70), enabling temporally coherent evaluations. Third, the model jointly downscales \textit{multiple atmospheric variables} (here, temperature, winds, and precipitation) using a flexible design to include more variables. Fourth, it provides a \textit{comprehensive evaluation}, including deterministic metrics, distributional fidelity, spectral recovery, and probabilistic skills. Our results show that the model accurately reconstructs fine-scale variability and extreme events, while its generative nature enables uncertainty quantification—essential for risk assessment. This study represents an initiative toward global to regional downscaling of climate projections: CMIP6, CMIP7, or other climate simulation efforts.}


\section{Methodology overview}\label{methodology_overview}
{
IPSL-AID is based on denoising diffusion probabilistic models (DDPMs). These models are trained to iteratively add and remove noise from data, thereby learning to reconstruct underlying fine-scale fields. For downscaling, we condition the model on coarse fields and their spatiotemporal context to generate multiple plausible realizations. While IPSL-AID encompasses the full design space of diffusion models introduced by \cite{karras_2022}, including variance-preserving, variance-expanding, and improved DDPM formulations, this work focuses on the Elucidated Diffusion Model (EDM). The core architecture is a U-Net implemented within the EDM framework. Appendix~\ref{Diffusion_model} provides a technical description of the model architecture, training procedure, loss function, sampling algorithm, and overall workflow.
}

\begin{figure}[t]
\FIG{\includegraphics[width=0.8\textwidth]{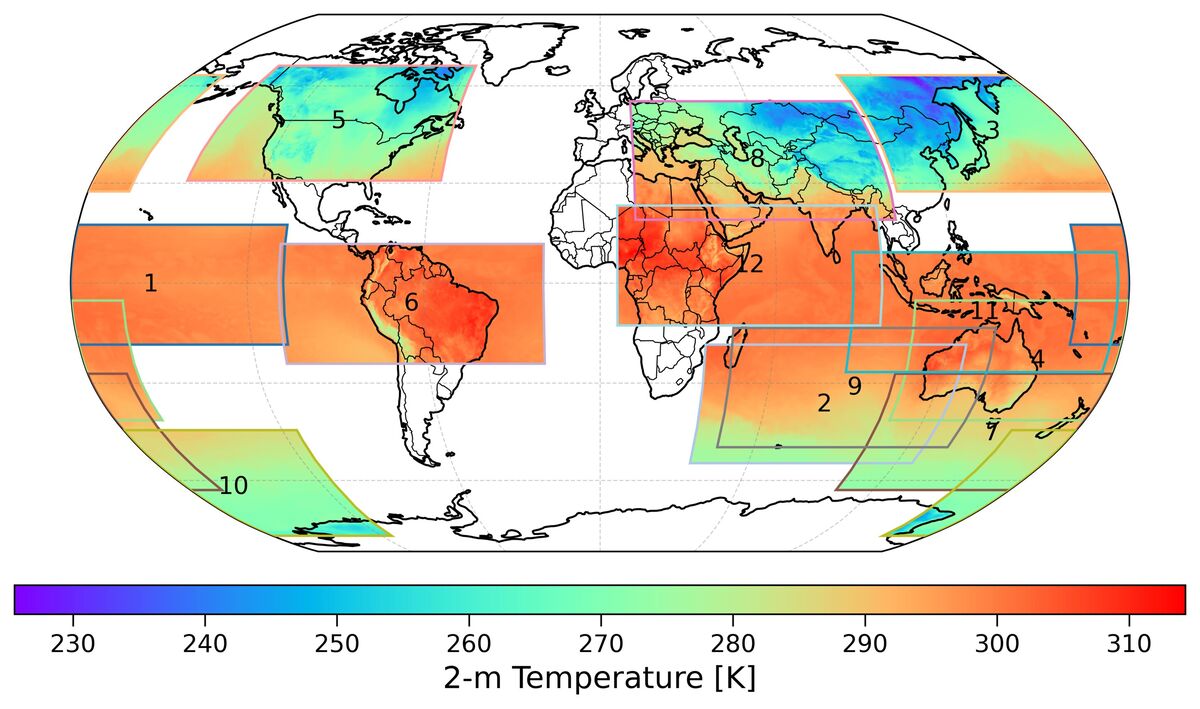}}
{\caption{An example of randomly sampled spatial blocks used during training of the global model}
\label{random-blocks-example}}
\end{figure}

\section{Data, sampling strategy, and metrics}\label{Data_sampling_metrics}

The model is trained using four ERA5 variables \citep{hersbach_era5_2020}: 2-m temperature (t2m), 10-m zonal wind (u10), 10-m meridional wind (v10), and precipitation (tp), at 0.25$^{\circ}$ resolution. To balance temporal variability and computational cost, we select four random time steps per day.

Training high-resolution (HR) global downscaling models is computationally intensive and requires substantial memory and GPU resources. To address this challenge, a random spatial block sampling strategy was developed. During each training epoch, $s$ spatial blocks of size $144\times360$ are generated, with block centers placed randomly (Figure \ref{random-blocks-example}). The longitude of each block is treated as periodic, while the latitude is constrained within valid global boundaries. Several values for the number of spatial blocks per epoch ($s=6$, 9, and 12) were tested, using $s=12$ was identified as an effective balance between computational efficiency and spatial diversity. For evaluation and inference, 20 fixed blocks corresponding to the ERA5 resolution of $1440\times721$ were used for global predictions. This framework is applicable to regional downscaling by specifying the center and spatial extent of the region.

A coarse-up procedure based on bilinear interpolation is used to separate large-scale and fine-scale components of the ERA5 fields. Let
$
\mathbf{y}^{\mathrm{HR}}
$ be a high-resolution (HR) field. The HR field is first reduced to a coarse resolution of $16\times32$ ({$\approx 2.25^\circ$ grid, comparable to typical GCMs resolution}) by bilinear interpolation, which eliminates small-scale variability while preserving large-scale structure. 
{ 
The resulting coarse field is then scaled back to the original HR grid using the same interpolation method, yielding a smooth coarse-up (CU) approximation, $\mathbf{y}^{\mathrm{CU}}$, unlike \citet{mardani_2025}, who typically rely on mean regression models.} The residual field, $\mathbf{R} = \mathbf{y}^{\mathrm{HR}} - \mathbf{y}^{\mathrm{CU}}$,
represents the fine-scale information lost during the upscaling process and serves as the training target for the diffusion model.

A conditional diffusion model is trained to generate fine-scale residual fields given CU fields and auxiliary spatiotemporal conditioning. The CU fields are provided as spatial conditioning inputs and are augmented with geographical variables, including latitude, longitude, topography ($z$), and land-sea mask (LSM). Temporal information is incorporated using cosine--sine representations of the day of year and hour of day (time embedding). The trained model generates residual fields ($\mathbf{R}'$) conditioned on the CU inputs and spatiotemporal context. The final HR prediction is obtained by adding the generated residual to the CU field ($\mathbf{y}^{\mathrm{CU}} + \mathbf{R}'$), which preserves large-scale consistency and enhances fine-scale details. {The workflow is shown in Appendix \ref{flow_chart}. We set the optimal parameters of the sampler following an ablation study (Appendix~\ref{ablation_study}). We trained a U-Net model as a baseline using the L2 loss function to predict the HR fields, using the network described in Section~\ref{architecture}. The inputs include the CU fields, static geographical variables, and temporal embeddings. The diffusion model has 92,140,548 trainable parameters, compared with 92,135,940 for the U-Net baseline.} The models were trained on data from 2015 to 2019, validated on 2020 data, and tested on 2021 data. Training ran for 100 epochs with a batch size of 70, using four NVIDIA A100 GPUs (64 GB each) on the LEONARDO supercomputer. Training, evaluation, and inference for the diffusion model took approximately 6 days.

\subsection{Metrics}
The model is evaluated using a set of metrics and spatial statistical diagnostics to assess both the pointwise accuracy and the distributional fidelity of the predictions. The primary metrics include the Mean Absolute Error $ \mathrm{MAE} = \frac{1}{N} \sum_{\mathrm n=1}^{\mathrm N} |y_{\mathrm n} - \hat{y}_{\mathrm n}|$, the Normalized MAE $ \mathrm{NMAE} = {\sum_{\mathrm n=1}^{\mathrm N} |y_{\mathrm n} - \hat{y}_{\mathrm n}|}/{\sum_{\mathrm n=1}^{\mathrm N} |y_{\mathrm n}|}$, the Root Mean Square Error $ \mathrm{RMSE} = \sqrt{\frac{1}{N} \sum_{\mathrm n=1}^{\mathrm N} (y_{\mathrm n} - \hat{y}_{\mathrm n})^2}$, the coefficient of determination $R^2 = 1 - {\sum_{\mathrm n=1}^{\mathrm N} (y_{\mathrm n} - \hat{y}_{\mathrm n})^2}/{\sum_{\mathrm n=1}^{\mathrm N} (y_{\mathrm n} - \bar{y})^2}$, and the Continuous Ranked Probability Score $\mathrm{CRPS}(F_{\mathrm{n}}, y_{\mathrm{n}}) = \int_{-\infty}^{\infty} \mathrm{d}\hat{y} \, (F_{\mathrm{n}}(\hat{y}) - \mathbf{1}(\hat{y} - y_{\mathrm{n}}))^2$. The CRPS is calculated using 100 time steps, for which the EDM sampler is run 10 times. Here, $y_{\mathrm n}$ and $\hat{y}_{\mathrm n}$ represent the HR fields from the ERA5 and predicted values, $\bar{y}$ is the mean of the truth, $N$ is the number of samples, and $F$ is the predictive cumulative distribution function. 

Power Spectral Density (PSD) analysis is used to evaluate the model's ability to reconstruct spatial variability across scales, from large-scale structures to smaller-scale features. Probability Density Functions (PDFs) on logarithmic scales enable comparisons of statistical distributions and the evaluation of extreme-value representation. Distributional consistency is further quantified using the Kullback--Leibler (KL) divergence $ \mathrm{KL}(p \parallel q) = \int {\mathrm d}y\, p(y) \log \frac{p(y)}{q(y)}$, where $p$ and $q$ denote the reference and predicted PDFs, respectively. Additionally, the Pearson correlation coefficient 
$\rho_{\mathrm{p},\mathrm{q}} = \frac{\mathrm{cov}(p,q)}{\sigma_{\mathrm{p}} \sigma_{\mathrm{q}}}$ is computed between the predicted and true fields to assess the similarity of their spatial patterns and overall structure, where $p$ and $q$ denote the flattened prediction and truth, respectively. The $q$th quantile of a sample $a$ is $Q(q) = (1-g)\,y[j] + g\,y[j+1]$, with $y$ sorted, $j = \lfloor (n-1)q \rfloor$, and $g = (n-1)q - j$.

{
For the diffusion model, we also evaluated rank histograms, which assess how well the predicted ensemble represents the truth, and the spread--skill ratio (SSR), which quantifies the consistency between ensemble spread and prediction error.
Given an ensemble of size $M$ (we used $M=10$), the sorted predictions are $\hat{y}_{(1)} \le \hat{y}_{(2)} \le \cdots \le \hat{y}_{(M)}$. The rank $r$ of reference $y_{\mathrm n}$ relative to the ensemble is given by $r = \sum_{\mathrm m=1}^{\mathrm M} \mathbf{1}(\hat{y}_{\mathrm m} \le y_{\mathrm n})$, 
so that $r \in \{0, \dots, M\}$. The rank histogram is obtained by plotting the histogram of $r$ over all prediction--truth pairs. The SSR is defined as 
$\mathrm{SSR} = \frac{\sigma_{\mathrm{ens}}}{\mathrm{RMSE}}$, where the ensemble spread is given by 
$\sigma_{\mathrm{ens}} = \sqrt{\frac{M+1}{M}} \sqrt{\frac{1}{N} \sum_{\mathrm n=1}^{\mathrm N} \frac{1}{M} \sum_{\mathrm m=1}^{\mathrm M} (\hat{y}_{\mathrm n}^{(\mathrm m)} - \bar{\hat{y}}_{\mathrm n} )^2}$, 
and 
$\mathrm{RMSE} = \sqrt{\frac{1}{N} \sum_{\mathrm n=1}^{\mathrm N} \left(\bar{\hat{y}}_{\mathrm n} - y_{\mathrm n} \right)^2}$. 
Here, $\hat{y}_{\mathrm n}^{(\mathrm m)}$ denotes the $m$-th ensemble member (with $m=1,\dots,M$), $\bar{\hat{y}}_{\mathrm n} = \frac{1}{M} \sum_{\mathrm m=1}^{\mathrm M} \hat{y}_{\mathrm n}^{(\mathrm m)}$ is the ensemble mean, $y_{\mathrm n}$ is the reference value, and $N$ is the number of samples. The corrective factor in the definition of $\sigma_{\mathrm{ens}}$ ensures the optimal value of the SSR is 1, as explained in \cite{fortin_2014}.

}

\section{Results and discussion}\label{Results_discussion}

\begin{figure}[t]%
\FIG{\includegraphics[width=0.95\textwidth]{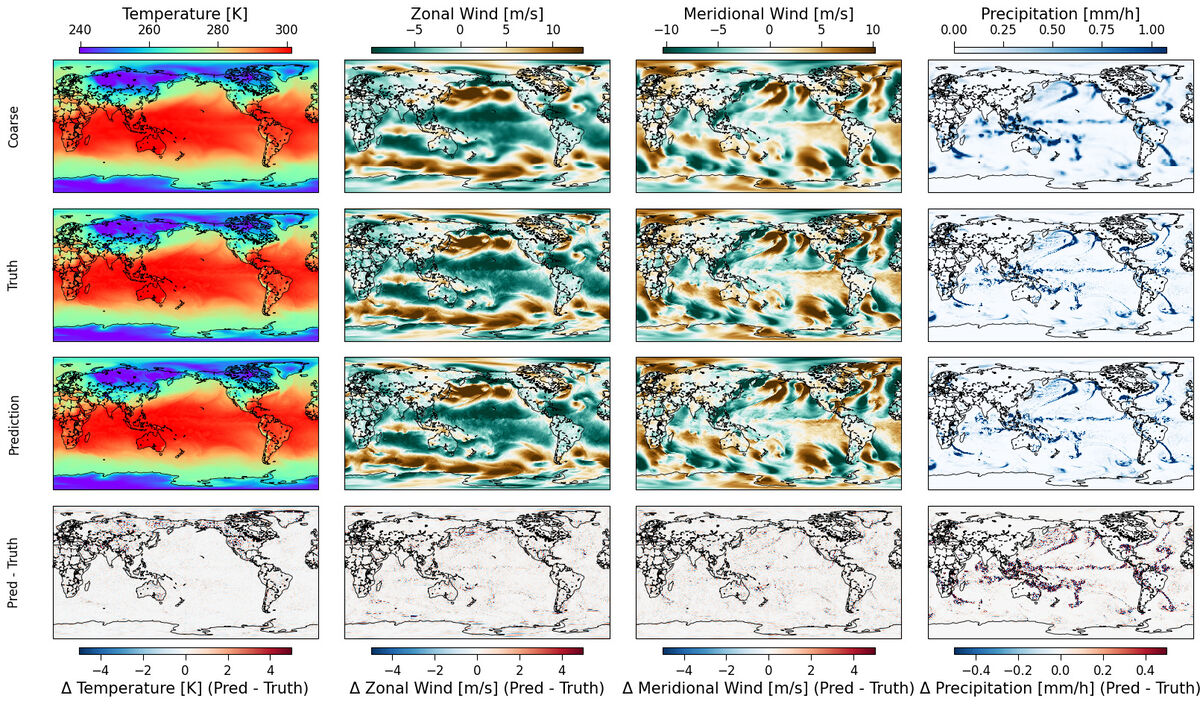}}
{\caption{Surface plots are shown for t2m, u10, v10, and tp (columns 1 to 4). The rows, from top to bottom, represent the CU ERA5 input, the HR ERA5 reference, the model prediction, and the difference between the prediction and the HR reference. All panels correspond to 2021-01-01 06:00 UTC}
\label{surface_plot_global}}
\end{figure}

Figure~\ref{surface_plot_global} shows surface plots for the downscaled variables: t2m, u10, v10, and tp. {Three representative members of a 10-member ensemble, the ensemble standard deviation, and a zoomed-in over South America are shown in Appendix~\ref{Ensemble_Mean} and Appendix~\ref{zoomed_in_example}, respectively}. Visually, the model successfully reconstructs the fine-scale spatial details of the HR fields. For t2m, the prediction achieves an MAE of $0.44 \pm 0.02$~K, an RMSE of $0.71 \pm 0.04$~K, and an $R^2$ of $1.00$. {The corresponding ensemble mean improves these scores, with an MAE of $0.32 \pm 0.01$~K, an RMSE of $0.52 \pm 0.01$~K, and an $R^2$ of $1.00$}. The wind components are predicted with similar accuracy. For u10, the prediction yields an MAE of $0.52 \pm 0.02$~m~s$^{-1}$, an RMSE of $0.79 \pm 0.03$~m~s$^{-1}$, and an $R^2$ of $0.98$, {while the ensemble mean achieves an MAE of $0.40 \pm 0.01$~m~s$^{-1}$, an RMSE of $0.62 \pm 0.03$~m~s$^{-1}$, and an $R^2$ of $0.988 \pm 0.002$}. Similarly, for v10, the prediction has an MAE of $0.50 \pm 0.01$~m~s$^{-1}$, an RMSE of $0.74 \pm 0.03$~m~s$^{-1}$, and an $R^2$ of $0.98$, {while the ensemble mean yields an MAE of $0.38 \pm 0.01$~m~s$^{-1}$, an RMSE of $0.59 \pm 0.03$~m~s$^{-1}$, and an $R^2$ of $0.985 \pm 0.002$}. For tp, the prediction shows an average MAE of $0.10 \pm 0.00$~mm~hr$^{-1}$, which appears very low; however, the NMAE of $0.67 \pm 0.02$ indicates a larger relative error. {The ensemble mean slightly improves the absolute error, with an MAE of $0.055 \pm 0.001$~mm~hr$^{-1}$ and an RMSE of $0.234 \pm 0.001$~mm~hr$^{-1}$, while the NMAE remains relatively high at $0.53 \pm 0.01$ and $R^2$ reaches $0.68 \pm 0.01$}. The discrepancy between MAE and NMAE arises from the variable's skewed distribution, characterized by a high frequency of zero values, which makes absolute error metrics less informative. Given its temporal and spatial variability, precipitation is one of the most complex fields to downscale; thus, the emphasis is on overall quality (see the following discussions).

The CRPS metric is $0.25 \pm 0.01$~K for t2m, $0.29 \pm 0.01$~m~s$^{-1}$ for u10, $0.28 \pm 0.01$~m~s$^{-1}$ for v10, and $0.04$~mm~hr$^{-1}$ for tp.
{ \citet{watt_2024} reported CRPS values of 0.254~K (t2m), 0.224~m~s$^{-1}$ (u10), and 0.232~m~s$^{-1}$ (v10) over the United States, whereas \citet{mardani_2025}, for the Taiwan domain, report higher values for extremes: 0.55~K, 0.86~m~s$^{-1}$, and 0.95~m~s$^{-1}$, respectively. The low CRPS for precipitation suggests that the model's predicted distribution is confident, particularly in correctly capturing the prevalence of dry (zero-precipitation) conditions. 
Overall, the diffusion model yields physically consistent (an inter-variable correlation analysis is presented in Appendix~\ref{correlation_analysis}) and statistically accurate downscaling, with the ensemble mean improving deterministic performance.}

\begin{figure}[t]%
\FIG{\includegraphics[width=0.9\textwidth]{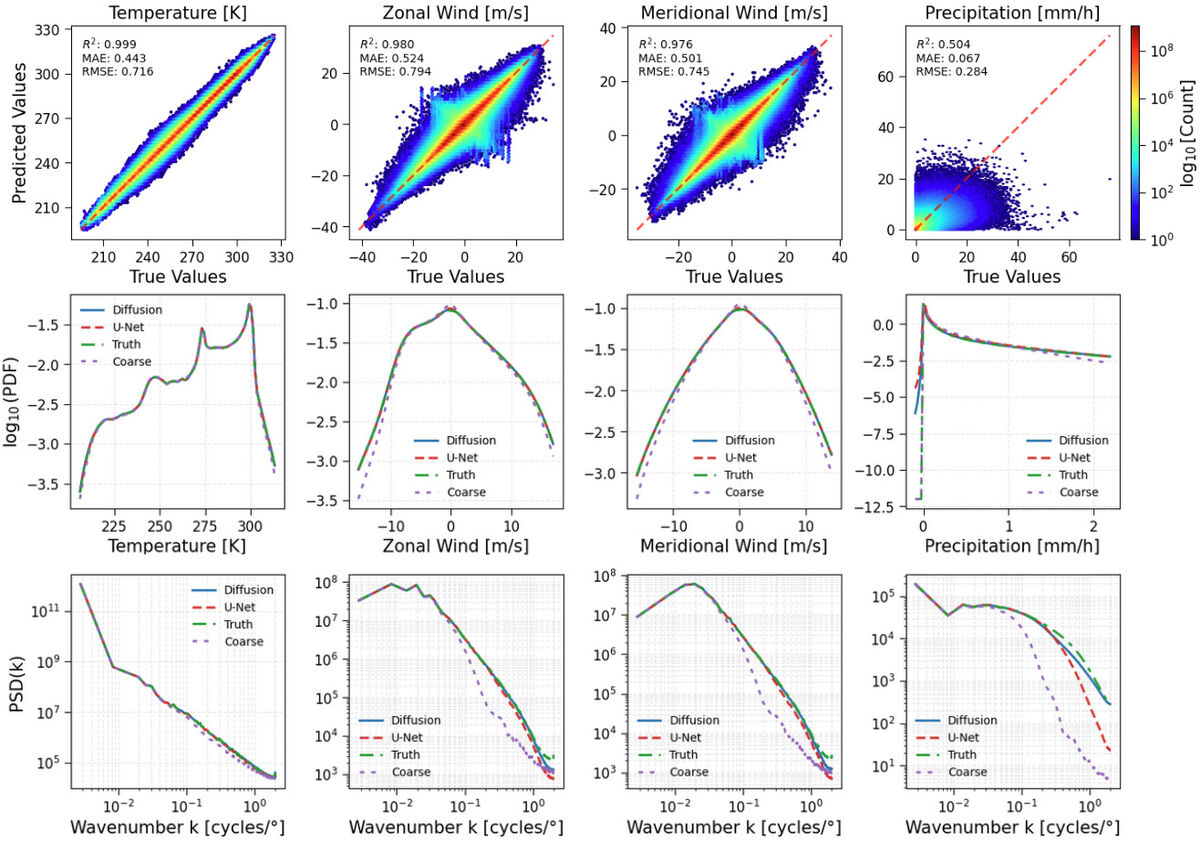}}
{\caption{Model evaluation for the complete 2021 evaluation dataset. Columns represent t2m, u10, v10, and tp, respectively. Rows display density scatter plots (top), PDFs (middle), and PSDs (bottom)}
\label{scatter_pdf_psd}}
\end{figure}

\begin{figure}[t]%
\FIG{\includegraphics[width=\textwidth]{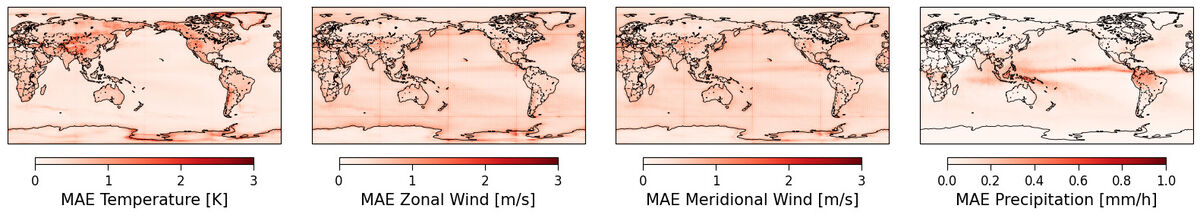}}
{\caption{Spatial distribution of the MAE, averaged over the 2021 dataset, for the downscaled variables: t2m, u10, v10, and tp}
\label{MAE_all_variables}}
\end{figure}

The statistical fidelity of the downscaled fields was evaluated by jointly analyzing density scatter plots, PDFs, and PSDs of the model predictions, the HR ERA5 reference, and the CU inputs (Figure~\ref{scatter_pdf_psd}). The first row displays density scatter plots comparing the diffusion model predictions with the HR reference across all variables. Each panel presents the joint distribution of values, with colors representing the logarithm of sample density and the one-to-one reference line. Quantitative metrics, including $R^2$, MAE, and RMSE, are also given. For the t2m and (u10, v10) components, the predictions show a strong linear relationship with the reference fields, as indicated by points tightly clustered along the diagonal and $R^2$ values near $0.99$. 
{Small vertical lines in the density plot for u10 and v10 indicate that some true values correspond to predicted values with the opposite sign. These discrepancies likely stem from challenges in downscaling coastal winds influenced by land–sea breezes (Figure \ref{MAE_all_variables}) and affect only a small fraction of datapoints.}
For precipitation, although the scatter indicates increased spread, particularly at higher intensities, the model substantially sharpens near the center, reduces underestimation, and more closely aligns with the one-to-one relationship, producing the same zero-precipitation points.

The model exhibits a strong capacity to replicate the full statistical distribution of t2m. The predicted mean ($\mu = 278.89$~K) and standard deviation ($\sigma = 21.33$~K) closely match those of the ground truth ($\mu = 278.91$~K, $\sigma = 21.32$~K), whereas the CU input displays a slightly reduced variance ($\sigma = 21.18$~K). This high level of distributional agreement is quantitatively supported by a KL divergence of $0.00$ and a Pearson correlation of $1.00$. The log-PDF curves of the prediction and reference are nearly indistinguishable across the t2m range, demonstrating accurate reproduction of both central tendencies and distribution tails. The model's performance for the (u10, v10) components is comparably robust. For the u10, both the mean ($\mu_{\mathrm{pred}} = -0.04$~m~s$^{-1}$, $\mu_{\mathrm{truth}} = -0.04$~m~s$^{-1}$) and variance ($\sigma_{\mathrm{pred}} = 5.63$~m~s$^{-1}$, $\sigma_{\mathrm{truth}} = 5.64$~m~s$^{-1}$) are accurately captured, while the CU input underestimates variability ($\sigma = 5.26$~m~s$^{-1}$). The predicted distribution aligns closely with the reference, as proved by a KL divergence of $0.00$ and a Pearson correlation of $0.99$. The v10 shows similar results, with the model recovering the true variance ($\sigma_{\mathrm{pred}} = 4.77$~m~s$^{-1}$, $\sigma_{\mathrm{truth}} = 4.78$~m~s$^{-1}$) from the smoothed CU field ($\sigma = 4.39$~m~s$^{-1}$), while maintaining a KL divergence of $0.00$ and a Pearson correlation of $0.99$. Precipitation prediction presents significant challenges due to its intermittency, non-Gaussian distribution, and localized extremes. Nevertheless, the model demonstrates substantial improvement over the CU input, with a KL divergence of $0.001$ and a Pearson correlation of $0.73$. The log-PDF indicates that the predictions more accurately capture the frequency of light-to-moderate precipitation and partially recover the tail of extreme events, which are suppressed in the CU field. 

{The U-Net baseline reproduces statistical distributions effectively (KL divergence and Pearson correlations of approximately $10^{-4}$ and 0.99, respectively). However, its performance declines for precipitation with a KL divergence of 0.3. Analysis of residual fields (Appendix~\ref{residual_fields}) shows that the U-Net generates PDFs that are more peaked and narrower than those from the diffusion model across variables, indicating underestimation of finescale variability and extremes.}

The model's ability to recover spatial variability across scales is evaluated by analyzing the PSDs of the generated fields and comparing them with the HR reference and the CU input, as illustrated in the third row of Figure~\ref{scatter_pdf_psd}. For all variables, the diffusion model prediction closely matches the reference spectrum throughout most of the resolved wavenumber range, extending into smaller-scale regimes. The CU input shows significant depletion at high wavenumbers, reflecting the loss of small-scale variability due to spatial averaging. In contrast, the predicted spectra maintain both the spectral slope and amplitude, indicating the model’s capacity to reproduce physically consistent small-scale variability. {U-Net spectra generally follow the large-scale truth data but exhibit earlier attenuation at higher wavenumbers, indicating a reduced ability to capture small-scale variability compared to the diffusion model}. In the case of precipitation, the recovery of high-wavenumber energy is especially pronounced, signifying the restoration of fine-scale structures associated with convective and localized rainfall processes, which remain parameterized at the 25 km resolution. Only minor deviations are observed at the highest wavenumbers, where all spectra are affected by numerical noise and resolution constraints.

\begin{table*}[t]
\tabcolsep=0pt%
\TBL{\caption{Extreme true, diffusion model, U-Net, and CU input quantiles for the downscaled variables.\label{extreme_qq}}}
{\begin{fntable}
\begin{tabular*}{\textwidth}{@{\extracolsep{\fill}}lcccc cccc cccc cccc@{}}\toprule%
 & \multicolumn{4}{@{}c@{}}{\TCH{t2m (K)}} 
 & \multicolumn{4}{@{}c@{}}{\TCH{u10 (m\,s$^{-1}$)}} 
 & \multicolumn{4}{@{}c@{}}{\TCH{v10 (m\,s$^{-1}$)}} 
 & \multicolumn{4}{@{}c@{}}{\TCH{tp (mm\,hr$^{-1}$)}} \\
\cmidrule{2-5}\cmidrule{6-9}\cmidrule{10-13}\cmidrule{14-17}%
\TCH{Quantile ($q$)} 
 & \TCH{Truth} & \TCH{Diff} & \TCH{U-Net} & \TCH{CU}
 & \TCH{Truth} & \TCH{Diff} & \TCH{U-Net} & \TCH{CU}
 & \TCH{Truth} & \TCH{Diff} & \TCH{U-Net} & \TCH{CU}
 & \TCH{Truth} & \TCH{Diff} & \TCH{U-Net} & \TCH{CU} \\\midrule

0.90  & 300.09 & 300.07 & 300.08 & 300.01 & 7.61 & 7.60 & 7.56 & 7.09 & 6.10 & 6.09 & 6.06 & 5.64 & 0.20 & 0.21 & 0.21 & 0.27 \\
0.95  & 301.20 & 301.17 & 301.14 & 300.97 & 10.24 & 10.24 & 10.19 & 9.71 & 7.93 & 7.91 & 7.89 & 7.31 & 0.47 & 0.49 & 0.49 & 0.49 \\
0.975 & 302.57 & 302.57 & 302.52 & 302.13 & 12.33 & 12.33 & 12.27 & 11.77 & 9.60 & 9.57 & 9.55 & 8.82 & 0.92 & 0.94 & 0.89 & 0.75 \\
0.99  & 306.40 & 306.41 & 306.35 & 305.55 & 14.48 & 14.47 & 14.40 & 13.85 & 11.62 & 11.58 & 11.56 & 10.66 & 1.71 & 1.73 & 1.56 & 1.13 \\
0.995 & 309.21 & 309.21 & 309.16 & 308.30 & 15.81 & 15.80 & 15.72 & 15.11 & 13.04 & 13.00 & 12.97 & 11.94 & 2.43 & 2.45 & 2.15 & 1.45 \\

\botrule
\end{tabular*}%
\end{fntable}}
\end{table*}

{The model's performance in predicting extreme values was quantified using quantile–quantile analysis of distribution upper tails (Table~\ref{extreme_qq}). For t2m, both diffusion model and the U-Net reconstruction closely match observed quantiles from 90th to 99.5th percentile. The diffusion model shows perfect agreement with truth (309.21 versus 309.21~K at $q=0.995$), while U-Net slightly underestimates the highest quantiles (309.16 versus 309.21~K). CU input underestimates high temperatures, with discrepancies reaching 0.91~K at the 99.5th percentile. For wind components, both models slightly underestimate extreme wind speeds. The diffusion model remains close to truth (15.80 versus 15.81~m~s$^{-1}$ at $q=0.995$ for u10), while U-Net shows larger underestimation (15.72~m~s$^{-1}$). Similar behavior occurs for v10. This shows both models reproduce most wind distribution, though capturing extreme events remains challenging. CU input performs worse, underestimating u10 extremes by 0.70~m~s$^{-1}$ at the 99.5th percentile. For precipitation, the diffusion model slightly underestimates moderate extremes but overestimates most extreme events (2.45 versus 2.43~mm~hr$^{-1}$ at $q=0.995$). The U-Net underestimates the upper tail more remarkably (2.15 versus 2.43~mm~hr$^{-1}$ at $q=0.995$). CU input overestimates lower precipitation quantiles but significantly underestimates the highest quantiles, with values below two-thirds of ground truth at $q=0.995$.}

Spatial error distributions indicate regional variation in model performance as illustrated in Figure~\ref{MAE_all_variables}. The MAE for t2m is higher at high latitudes, particularly near the poles, and in areas with complex topography, such as the Himalayas and the Andes. The (u10, v10) components also show increased errors in coastal regions and along steep topographic gradients, likely attributable to the modulation of surface winds by land-sea contrasts and orographic influences. In contrast, precipitation errors are most pronounced in the tropics, especially within the Intertropical Convergence Zone and over tropical rainforests, where intense and localized rainfall events pose significant challenges for downscaling. { The analogous maps illustrating the relative error are provided in Appendix~\ref{relative_errors}.}

{
The rank histograms shown in Figure~\ref{rank_histograms} are generally close to uniform, indicating that the diffusion model ensemble is reasonably well calibrated overall. However, the temperature histogram exhibits a slight upward slope with a negative bias, suggesting that the truth often lies above the ensemble members, indicating a tendency toward underestimation. The u10 histogram is fairly flat but shows a mild positive bias, pointing to a slight overestimation. In contrast, the v10 histogram is nearly uniform with negligible bias, indicating very good calibration. The precipitation histogram displays a noticeable spike at the lowest rank and a slight skew, suggesting that the truth frequently falls below the ensemble range, which implies overestimation and potential under-dispersion. 

\begin{figure}[b]%
\FIG{\includegraphics[width=\textwidth]{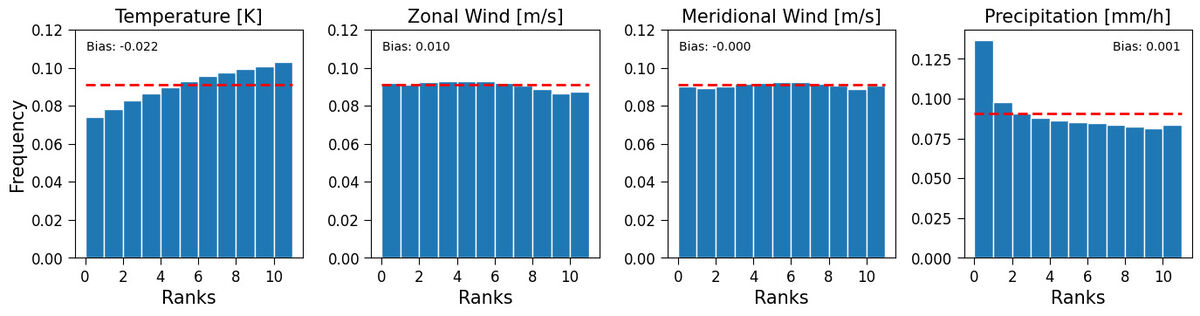}}
{\caption{Rank histograms for the 10-member diffusion ensemble. Columns show t2m, u10, v10, and tp. Red dashed lines mark the expected uniform frequency}
\label{rank_histograms}}
\end{figure}

Figure~\ref{ssr_hexbin} presents density plots of ensemble spread versus RMSE. The red dashed line represents the ideal 1:1 relationship (SSR = 1), which corresponds to a perfectly calibrated ensemble. All variables display SSR values slightly below 1, indicating a mild overall under-dispersion. The highest sample density is concentrated at low RMSE and low spread for all variables, which reflects strong predictive skill. Deviations from the 1:1 line become increasingly evident as the RMSE values increase. For u10, v10, and tp, the spread remains relatively small as RMSE increases, suggesting under-dispersion in higher-error regimes. t2m exhibits a similar, though less pronounced, pattern.

\begin{figure}[t]%
\FIG{\includegraphics[width=\textwidth]{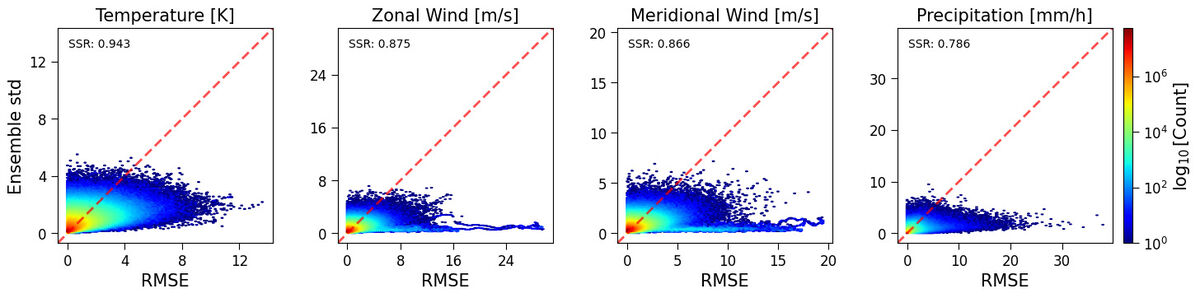}}
{\caption{Density plots of ensemble spread versus RMSE for t2m, u10, v10, and tp. The red dashed line indicates the 1:1 relationship corresponding to perfect spread--skill (SSR = 1)}
\label{ssr_hexbin}}
\end{figure}
}
\section{Conclusions and limitations}
{ 
This study introduces IPSL-AID, a diffusion-based generative model developed for downscaling climate data from global to regional scales. IPSL-AID reconstructs high-resolution fields of temperature, winds, and precipitation from coarse input data and their spatiotemporal context. The model is trained on ERA5 reanalysis using a novel random block sampling strategy, which allows for global applicability without the need for retraining in new regions. A key advantage of this approach is its probabilistic framework. By learning the underlying distribution of residuals at fine scales, IPSL-AID generates multiple plausible high-resolution realizations from a single coarse input. This capability enables robust uncertainty quantification through ensemble spread analysis, rank histograms, and spread–skill ratio diagnostics.

Quantitative evaluation indicates that IPSL-AID achieves low deterministic errors (MAE of 0.44 K for T2m, 0.52 m s$^{-1}$ for u10, and 0.50 m s$^{-1}$ for v10) and accurately represents the statistical distribution of the variables. The model reproduces spatial variability across scales and captures extreme events, with quantile estimates closely matching observations up to the 99.5th percentile. IPSL-AID demonstrates robust performance across diverse atmospheric variables and regions while maintaining physical consistency, as shown by inter-variable correlation analysis that reveals realistic multivariate dependencies, such as temperature–wind patterns over oceans and continents. The ensemble mean further enhances deterministic scores, reducing MAE to 0.32 K for T2m and 0.40 m s$^{-1}$ for u10.

Despite its strengths, IPSL-AID exhibits certain limitations. Performance degrades in regions with complex topography (e.g., Himalayas, Andes), coastal zones, and areas characterized by intense convective precipitation, where errors remain elevated. For precipitation, the normalized MAE of 0.53 indicates a substantial relative error, despite a low absolute error, reflecting the inherent challenge of downscaling non-Gaussian fields. The rank histograms reveal systematic tendencies: a slight underestimation for temperature, a mild overestimation for u10 winds, and under-dispersion for precipitation extremes. Spread–skill analysis indicates under-dispersion in higher-error regimes, suggesting the ensemble may not fully capture uncertainty in challenging conditions. Additionally, block boundary artifacts remain visible during global inference, and the five-year training period is short compared to the multi-decadal timescales typically required for climate studies. The computational demands of global training—6 days on four A100 GPUs—may also limit accessibility for some research groups.

Several avenues for future development are planned. First, we aim to extend the model to additional atmospheric variables and apply the framework to downscale future projections from CMIP6 models. This will require developing and integrating appropriate bias correction or debiasing strategies \citep{wan_2023, wan_2025}. To address computational limitations, distributed multi-GPU training strategies will be explored to accommodate longer training periods (e.g., 40 years) while maintaining global coverage. For block boundary artifacts, several mitigation approaches are under investigation: (i) an overlap-block method with cosine-weighted blending during inference, (ii) explicit boundary regularization in the loss function to penalize discontinuities along seam lines, (iii) frequency-domain spectral loss to suppress characteristic artifact frequencies, and (iv) adaptive Fourier-domain filtering as a post-processing step.

Finally, we plan to incorporate additional conditioning variables (e.g., soil moisture, surface fluxes) and explore alternative diffusion formulations to improve performance for precipitation extremes. The apparent contradiction between low CRPS and high NMAE for precipitation reflects the different aspects they measure: CRPS captures probabilistic calibration, whereas NMAE reflects the difficulty of deterministic prediction for a non-Gaussian variable; with conditioning limited to topography, strong deterministic skill is not achieved.
}
\begin{Backmatter}

\paragraph{Acknowledgments}
We acknowledge the EuroHPC Joint Undertaking for awarding this project access to the EuroHPC supercomputer LEONARDO, hosted by CINECA (Italy) and the LEONARDO consortium through an EuroHPC Development Access call (EHPC-DEV-2024D10-070, EHPC-DEV-2025D08-095).

\paragraph{Funding Statement}
Project S24JROI011 supported this work from the Grantham Foundation (To K.Ardaneh and O. Boucher). R. Lguensat acknowledges support from Agence Nationale de la Recherche - France 2030 through the PEPR TRACCS programme under grant number ANR-22-EXTR-0011 (LOCALISING).

\paragraph{Competing Interests}
None.

\paragraph{Data Availability Statement}
The IPSL-AID source code is publicly available on GitHub at \url{https://github.com/kardaneh/IPSL-AID}. The specific code version, along with the pre-trained model checkpoints, configuration scripts, and the 2021 ERA5 dataset used for inference in this study, are archived and publicly accessible on Zenodo at \url{https://doi.org/10.5281/zenodo.19200763}.

\paragraph{Ethical Standards}
The research meets all ethical guidelines, including adherence to the legal requirements of the study country.

\paragraph{Author Contributions}
Conceptualization: All authors contributed equally.  
Data curation: Kishanthan Kingston;  
Formal analysis: All authors contributed equally.  
Funding acquisition: Jean-Francois Lamarque, Olivier Boucher, Kazem Ardaneh;  
Investigation: All authors contributed equally.  
Methodology: All authors contributed equally.  
Project administration: Olivier Boucher, Kazem Ardaneh;  
Resources: Olivier Boucher, Kazem Ardaneh;  
Software: Kazem Ardaneh, Kishanthan Kingston;  
Supervision: Jean-Francois Lamarque, Olivier Boucher, Freddy Bouchet;  
Validation: Kishanthan Kingston, Kazem Ardaneh, Pierre Chapel;  
Visualization: Kishanthan Kingston, Kazem Ardaneh, Pierre Chapel;  
Writing – original draft: Kazem Ardaneh;  
Writing – review and editing: All authors contributed equally. All authors have reviewed and approved the final manuscript.

\begin{appendix}\appheader

\section{Generative diffusion model}\label{Diffusion_model}
\renewcommand{\thefigure}{A\arabic{figure}}  
\setcounter{figure}{0}  
\renewcommand{\thetable}{A\arabic{table}}
\setcounter{table}{0}
\cite{karras_2022} introduced a unified design space that categorizes the architectural and procedural choices of diffusion-based generative models. While IPSL-AID encompasses the full design space, including elucidated diffusion model (EDM), variance-preserving, variance-expanding, and improved DDPM, this work focuses on the EDM and reports its results.

\subsection{Preconditioning}
The preconditioned denoising model stabilizes training by standardizing the scales of inputs, outputs, and targets across varying noise levels. The EDM preconditioned model $D_\theta(\mathbf{x}; \sigma)$ is defined as follows:

\begin{equation}
D_\theta(\mathbf{x}; \sigma) =  c_{\mathrm {skip}}(\sigma) \mathbf{x} + c_{\mathrm{out}}(\sigma) F_\theta\big(c_{\mathrm{in}}(\sigma) \mathbf{x}; c_\mathrm{noise}(\sigma)\big)
\end{equation}
The input $\mathbf{x}$ is defined as $\mathbf{x}=\mathbf{y}+\sigma\mathbf{n}$, where $\mathbf{y}$ is the clean signal and $\mathbf{n} \sim \mathcal{N}(\mathbf{0}, \mathbf{1})$ is standard Gaussian noise. Here, $\sigma$ is the noise level, $F_\theta$ is the underlying neural network, and the coefficients 
$c_\mathrm{in}$ (that ensure the network sees inputs with roughly unit variance across all noise levels), $c_\mathrm{out}$ (that ensure that the network output has the correct magnitude to match the target), $c_\mathrm{skip}$ (that pass part of the input directly to the output to reduce the amplification of network errors), and the noise embedding $c_\mathrm{noise}$ (that gives the network explicit information about the noise level, allowing it to adapt its denoising behavior) are
$c_\mathrm{in}(\sigma) = 1 / (\sigma_\mathrm{data}^2 + \sigma^2)^{1/2}$, 
$c_\mathrm{skip}(\sigma) = \sigma_\mathrm{data}^2 / (\sigma_\mathrm{data}^2 + \sigma^2)$, 
$c_\mathrm{out}(\sigma) = \sigma \, \sigma_\mathrm{data} / (\sigma_\mathrm{data}^2 + \sigma^2)^{1/2}$, and $c_\mathrm{noise}(\sigma) =1/4 \log \sigma$, respectively, with $\sigma_{\mathrm{data}}$ is the standard deviation of the normalized data distribution, typically set to $\sigma_{\mathrm{data}} = 1$. The noise scale $\sigma$ is sampled from a log-normal distribution, $\log \sigma \sim \mathcal{N}(P_{\text{mean}}, P_{\text{std}}^2)$. The target for the network is preconditioned as:
\begin{equation}
    F_\text{target}(\mathbf{y},\mathbf{n};\sigma) = \frac{\mathbf{y} - c_\text{skip}(\sigma) \mathbf{x}}{c_\text{out}(\sigma)}
\end{equation}

{For conditional image generation, we extended the preconditioning framework by concatenating the conditioning image with the noised input along the channel dimension before passing it to the network.}
\subsection{Architecture}\label{architecture}
The EDM preconditioning architecture employs a U-Net-based denoising network, adapted from the work of \cite{dhariwal_2021}. It utilizes an encoder-decoder structure with symmetric downsampling and upsampling paths, incorporating multi-head self-attention blocks at selected spatial resolutions. The input consists of a noisy image $\mathbf{x} \in \mathbb{R}^{C \times H \times W}$, a scalar noise level or timestep embedding $\sigma$, and optional class or augmentation embeddings. {The output is a denoised tensor $\mathbf{y}'$.}

\subsubsection{Default Configuration}
The U-Net configuration includes a base channel count of $C_\mathrm{base} = 128$, channel multipliers per resolution of $[1, 2, 3, 4]$, and three residual blocks per resolution. Self-attention mechanisms are implemented at resolutions $[32, 16, 8]$ with a dropout probability of $p = 0.10$. The embedding dimension multiplier is defined as $C_\mathrm{emb} = 4 \times C_\mathrm{base}$. All convolutional and linear layers are initialized using Kaiming uniform initialization. The network architecture accommodates optional class-conditioning and augmentation embeddings, which are incorporated via linear projections applied to the noise embeddings.

\subsubsection{Mapping and Embedding Layers}
Noise levels $\sigma$ are represented using a sinusoidal positional embedding layer:
$\mathbf{e}_\sigma = \text{PE}(\sigma) \in \mathbb{R}^{C_\mathrm{base}}$
This representation is processed by two fully connected layers with SiLU activations: $\mathbf{e} = \text{SiLU}(W_1 \mathbf{e}_\sigma + b_1)$ and $\mathbf{e}= \text{SiLU}(W_2 \mathbf{e} + b_2)$. Optional class embeddings $\mathbf{y}_\text{class}$ and augmentation embeddings $\mathbf{y}_\text{aug}$ are projected into the same embedding space and added to $\mathbf{e}$ to provide conditioning information to the network. All linear layers are initialized using Kaiming uniform initialization, except for the class embedding projection, which uses Kaiming normal initialization. {For the U-Net network to serve as a deterministic baseline, the noise-embedding components are bypassed, and the embedding is constructed exclusively from the class and augmentation labels.}

\subsubsection{Encoder, Decoder, and Attention}
The encoder consists of convolutional and residual (\texttt{U-NetBlock}) layers, which progressively downsample the feature maps. At level $l$, the feature map has height
$H_{\mathrm l} = \lfloor H / 2^{\mathrm l} \rfloor$ and width $W_{\mathrm l} = \lfloor W / 2^{\mathrm l} \rfloor$. Level 0 starts with a $3 \times 3$ convolution to expand channels, while levels $l>0$ begin with a downsampling \texttt{U-NetBlock}. Each level then applies $N_\mathrm{block}$ residual \texttt{U-NetBlock} layers, updating the feature map as
$\mathbf{h}_{\mathrm{l,k+1}} = \text{U-NetBlock}(\mathbf{h}_{\mathrm{l,k}}, \mathbf{e})$, where $\mathbf{h}_{\mathrm{l,k}} \in \mathbb{R}^{C \times H_{\mathrm l} \times W_{\mathrm l}}$ and $\mathbf{e}$ denotes the combined embedding. Each \texttt{U-NetBlock} incorporates group normalization, two $3 \times 3$ convolutions, adaptive scaling with $\mathbf{e}$, and optional self-attention. Skip connections $\mathbf{s}_{\mathrm l}$ are retained for subsequent use in the decoder.

The decoder architecture mirrors the encoder's architecture. At each level, the corresponding skip connection from the encoder is concatenated with the decoder feature map.
$\mathbf{h}_{\mathrm{l-1,0}} = \text{Concat}[\mathbf{h}_{\mathrm{l,k}}, \mathbf{s}_{\mathrm l}]$, which is then processed by upsampling and $N_\mathrm{block}+1$ \texttt{U-NetBlock} layers. At the coarsest level, two \texttt{U-NetBlock} layers are applied before upsampling. Channel counts are determined by the base channel number and per-level multipliers. The final output is $\mathbf{y} = \text{Conv}_{3\times3}(\text{SiLU}(\text{GroupNorm}(\mathbf{h}_0)))$.

Self-attention is applied at spatial resolutions $[32, 16, 8]$, with 64 channels allocated per attention head. The attention operation is defined as follows:
$
\text{Attention}(\mathbf{Q}, \mathbf{K}, \mathbf{V}) = \text{softmax}({\mathbf{Q}^\top \mathbf{K}}/{\sqrt{d_{\mathrm k}}})\mathbf{V}
$.

\begin{figure}[t]
 \centering
 \FIG{\includegraphics[width=0.95\textwidth]{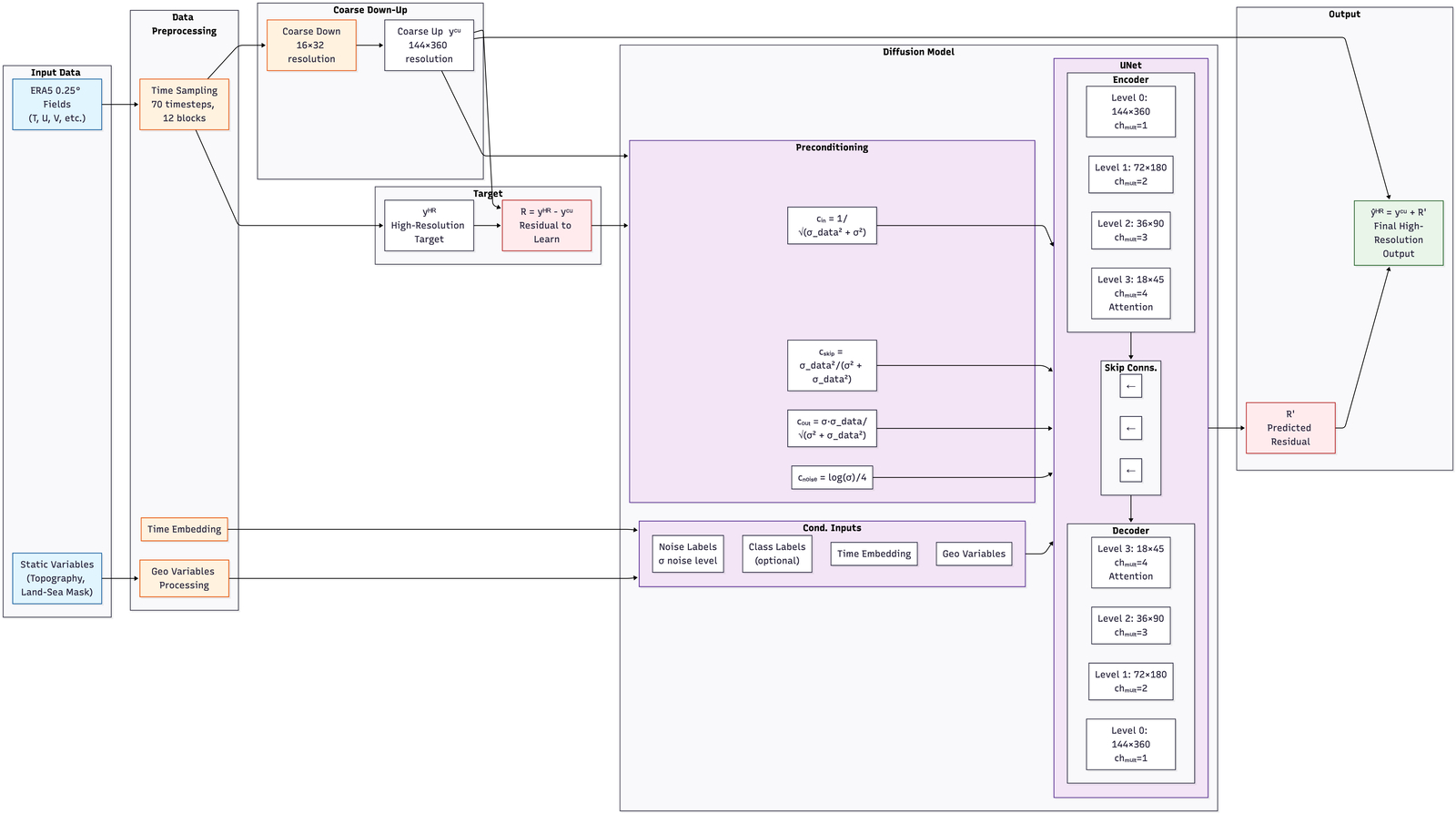}}
 {\caption{Data flow and model architecture for the conditional EDM model}
 \label{fig:supp1}}
\end{figure}

\subsection{Loss Function}\label{Loss Function}
For each training sample, Gaussian noise $\sigma \mathbf{n}$, with a randomly selected noise level $\sigma$, is added to the image. The network $D_{\theta}$ is then trained to predict the original clean image $\mathbf{y}$. The loss function is weighted by $\lambda(\sigma)$, which balances the contributions of different noise levels to the total loss and is defined as $\lambda(\sigma) = (\sigma^2 + \sigma_{\mathrm{data}}^2) / (\sigma \, \sigma_{\mathrm{data}})^2$. The weighted denoising loss reads:

\begin{equation}
\mathcal{L} =
\mathbb{E}_{\sigma, \mathbf{y}, \mathbf{n}}
\left[
\lambda(\sigma)
\left\|
D_{\theta}(\mathbf{y} + \sigma\mathbf{n}, \sigma)
- \mathbf{y}
\right\|_2^2
\right]
\label{eq:edm_loss}
\end{equation}

{
The loss function is augmented with a conditional image fed into the network, so that the model's predictions—and thus the loss—are calculated based on both the noisy image and the conditional input.
}

\subsection{Sampler}
The samples are generated from a trained diffusion model by numerically integerating the reverse-time stochastic differential equation. 
Let $N$ be the number of integrating steps, and $\sigma_{\min}$, $\sigma_{\max}$ be the smallest and largest noise scales, respectively. The sequence of time steps $t_{\mathrm{i}}$ is defined as
$t_{\mathrm{i}} = (\sigma_{\max}^{1/\rho} + \frac{\mathrm{i}}{N-1} (\sigma_{\min}^{1/\rho} - \sigma_{\max}^{1/\rho}) )^\rho$ for $\mathrm{i} = 0, \dots, N-1$, with $t_{\mathrm N} = 0$, 
where $\rho$ controls the step distribution (higher $\rho$ increases the density of steps at low noise levels). The sample is initialized with Gaussian noise $\mathbf{x}_0 \sim \mathcal{N}(\mathbf{0}, t_0^2 \mathbf{1})$. At each step, the latent is updated using the 1st- or 2nd-order Euler/Heun procedure as follows:

\begin{enumerate}[1.]
\item{
A temporary noise increment $\gamma_{\mathrm{i}}$ may be added if $t_{\mathrm{i}}$ falls within a specified range $[S_{\min}, S_{\max}]$, defined as $\gamma_{\mathrm{i}} = \min(S_{\mathrm{churn}}/N, \sqrt{2}-1)$ if $t_{\mathrm{i}} \in [S_{\min}, S_{\max}]$, and $\gamma_{\mathrm{i}} = 0$ otherwise. The temporarily increased noise level is then $\hat{t}_{\mathrm{i}} = t_{\mathrm{i}} + \gamma_{\mathrm{i}} \, t_{\mathrm{i}}$.
}

\item{Add new Gaussian noise to move from $t_{\mathrm i}$ to $\hat{t}_{\mathrm i}$, 
$\hat{\mathbf{x}}_{\mathrm i} = \mathbf{x}_{\mathrm i} + \sqrt{\hat{t}_{\mathrm i}^2 - t_{\mathrm i}^2} \, \boldsymbol{\epsilon}_{\mathrm i}$ where $\boldsymbol{\epsilon}_{\mathrm i} \sim \mathcal{N}(\mathbf{0}, S_{\mathrm noise} \mathbf{1})$.}

\item{Compute the first-order derivative of the reverse-time SDE as 
$\mathbf{d}_{\mathrm{i}} = (\hat{\mathbf{x}}_{\mathrm{i}} - D_\theta(\hat{\mathbf{x}}_{\mathrm{i}}, \hat{t}_{\mathrm{i}})) / \hat{t}_{\mathrm{i}}$, 
and update the latent using a forward Euler step 
$\mathbf{x}_{\mathrm{i+1}} = \hat{\mathbf{x}}_{\mathrm{i}} + (t_{\mathrm{i+1}} - \hat{t}_{\mathrm{i}}) \, \mathbf{d}_{\mathrm{i}}$.
}

\item{If $t_{\mathrm{i+1}} \neq 0$, apply the 2nd-order correction by first computing 
$\mathbf{d}_{\mathrm{i}}' = (\mathbf{x}_{\mathrm{i+1}} - D_\theta(\mathbf{x}_{\mathrm{i+1}}, t_{\mathrm{i+1}})) / t_{\mathrm{i+1}}$, 
and then updating the latent as 
$\mathbf{x}_{\mathrm{i+1}} = \hat{\mathbf{x}}_{\mathrm{i}} + (t_{\mathrm{i+1}} - \hat{t}_{\mathrm{i}}) \, \frac{1}{2} (\mathbf{d}_{\mathrm{i}} + \mathbf{d}_{\mathrm{i}}')$. } 

\item{Return the final denoised sample $\mathbf{x}_{\mathrm N}$.}

\end{enumerate}

\subsection{Workflow}\label{flow_chart}
{
Figure~\ref{fig:supp1} illustrates the data flow and model architecture for generating HR outputs from ERA5 data. The ERA5 data at 0.25$^{\circ}$ resolution ($\mathbf{y}^{\mathrm{HR}}$) are first sampled into sequences of 70 time steps with 12 blocks. Geospatial variables and temporal embeddings are extracted and combined with the coarse-upsampled $\mathbf{y}^{\mathrm{CU}}$ and the residual $\mathbf{R} = \mathbf{y}^{\mathrm{HR}} - \mathbf{y}^{\mathrm{CU}}$ as inputs to the model. The model predicts the residual $\mathbf{R}'$, which is added back to the CU to produce the final HR prediction $\mathbf{\hat{y}}^{\mathrm{HR}}$.
}

\begin{table*}[t]
\tabcolsep=0pt%
\TBL{\caption{Ablation study of EDM sampler parameters.\label{tab:edm_ablation}}}
{\begin{fntable}
\begin{tabular*}{\textwidth}{@{\extracolsep{\fill}}l ccc ccc ccc ccc @{}}\toprule%
 & \multicolumn{3}{c}{\TCH{$N$}} 
 & \multicolumn{3}{c}{\TCH{$\sigma_{\min}$}} 
 & \multicolumn{3}{c}{\TCH{$\sigma_{\max}$}} 
 & \multicolumn{3}{c}{\TCH{$\rho$}} \\
\cmidrule{2-4}\cmidrule{5-7}\cmidrule{8-10}\cmidrule{11-13}%
\TCH{Metric} 
 & \TCH{10} & \TCH{20} & \TCH{40}
 & \TCH{0.002} & \TCH{0.02} & \TCH{0.2}
 & \TCH{8} & \TCH{80} & \TCH{100}
 & \TCH{-10} & \TCH{7} & \TCH{10} \\
\midrule
MAE (K) 
 & 0.63 & 0.55 & 0.53
 & 0.63 & 0.58 & 0.46
 & 0.55 & 0.63 & 0.64
 & 0.64 & 0.63 & 0.61 \\
RMSE (K) 
 & 0.99 & 0.89 & 0.85
 & 0.99 & 0.93 & 0.76
 & 0.87 & 0.99 & 1.00
 & 1.04 & 0.99 & 0.96 \\
$R^2$ 
 & 0.998 & 0.998 & 0.998
 & 0.998 & 0.998 & 0.999
 & 0.998 & 0.998 & 0.998
 & 0.998 & 0.998 & 0.998 \\
Time 
 & 2h38 & 3h43 & 6h05
 & 2h38 & 2h33 & 2h35
 & 2h33 & 2h38 & 2h32
 & 2h32 & 2h38 & 2h38 \\
\botrule
\end{tabular*}%
\end{fntable}}
\end{table*}

\section{Extended diagnostics}
\renewcommand{\thefigure}{B\arabic{figure}}  
\setcounter{figure}{0}  
\renewcommand{\thetable}{B\arabic{table}}
\setcounter{table}{0}
{

\subsection{Ablation study}\label{ablation_study}
An ablation study was conducted to evaluate the impact of the EDM sampler parameters: the number of sampling steps ($N$), minimum noise level ($\sigma_{\min}$), maximum noise level ($\sigma_{\max}$), and time step exponent ($\rho$), on reconstruction accuracy and computational cost. The study was performed for t2m over 10 training epochs. The objective was to identify a configuration that balances prediction quality with inference efficiency. The baseline configuration was $N=10$, $\sigma_{\min}=0.002$, $\sigma_{\max}=80$, and $\rho=7$. As shown in Table~\ref{tab:edm_ablation}, increasing $N$ from 10 to 40 improves performance, reducing MAE from 0.63~K to 0.53~K and RMSE from 0.99~K to 0.85~K. However, inference time increases more than doubles (2h38 to 6h05). The largest performance improvement arises from increasing $\sigma_{\min}$. Raising it from 0.002 to 0.2 yields the best results, with MAE of 0.46~K and RMSE of 0.76~K, while maintaining a similar runtime (2h35). This suggests that a higher minimum noise level helps prevent overfitting to high-frequency details. In contrast, $\sigma_{\max}$ and $\rho$ have smaller effects. A moderate $\sigma_{\max}$ of 8 slightly improves performance (MAE: 0.55~K) compared to the default value of 80. The exponent $\rho$ has only minor influence, with $\rho=10$ marginally outperforming $\rho=7$, while $\rho=-10$ gives the worst results.

\subsection{Ensemble Mean}\label{Ensemble_Mean}
We utilize a 10-member ensemble to quantify plausible HR realizations and assess uncertainty in the diffusion model predictions. Figure~\ref{ensemble_members} shows surface plots for t2m, u10, v10, and tp (columns 1--4). The first three rows display three representative ensemble members, illustrating the variability among predictions, while the fourth row presents the ensemble mean, averaged over 10-members, which smooths small-scale variability and yields a more deterministic pattern.  The last row presents the ensemble standard deviation, highlighting regions of higher uncertainty.

\begin{figure}[t]%
\FIG{\includegraphics[width=0.9\textwidth]{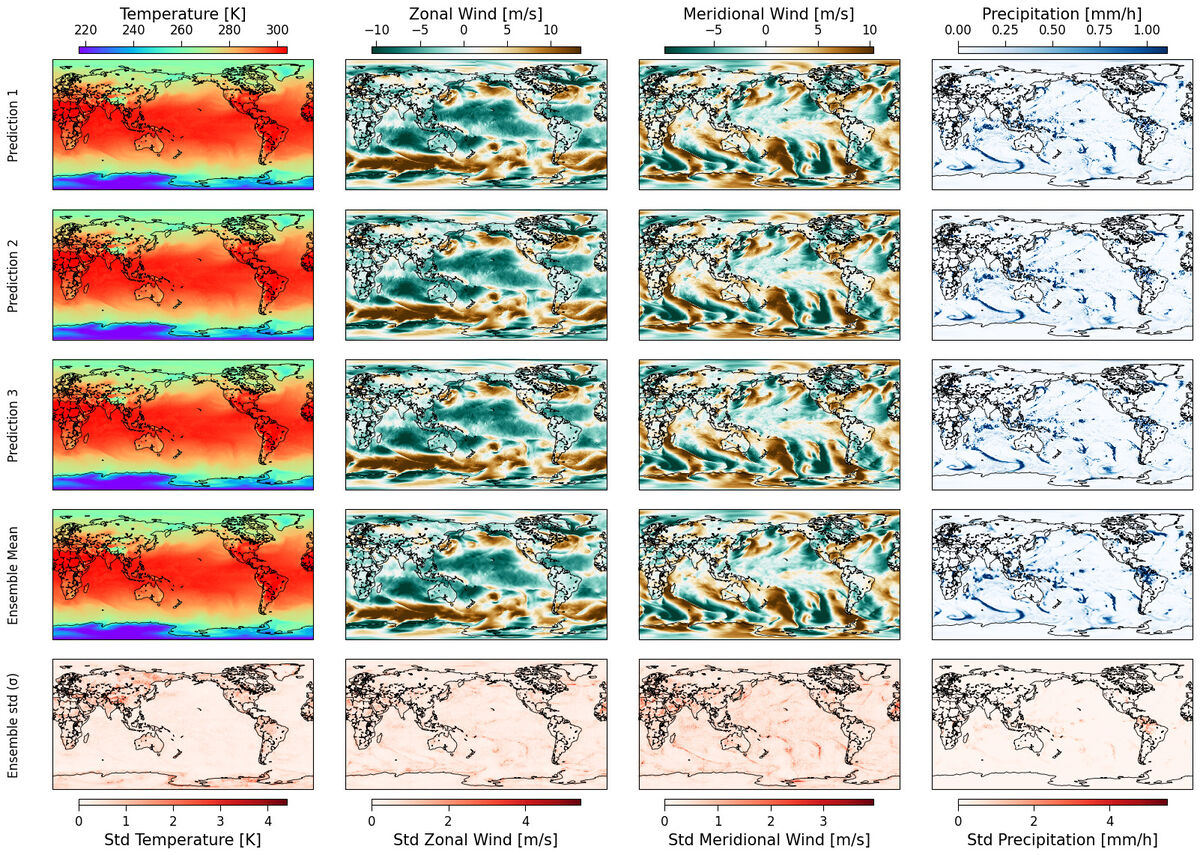}}
{\caption{Surface plots for t2m, u10, v10, and tp (columns 1 to 4). The first three rows show three representative members from the 10-member ensemble. The last row shows the ensemble standard deviation. All panels correspond to 2021-05-02 19:00 UTC}
\label{ensemble_members}}
\end{figure}

\subsection{Zoomed-in example}\label{zoomed_in_example}
Figure~\ref{zoomin} provides a zoomed-in comparison of U-Net and diffusion model predictions against the truth over South America. Whereas the U-Net accurately reproduces large-scale patterns, it clearly over-smooths fine spatial details for precipitation and winds. By contrast, the diffusion model distinctly reproduces high-frequency structures. Residual errors, as indicated by MAE, are most pronounced over the Andes and in regions of intense convective activity, such as the Amazon.

\begin{figure}[htb!]%
\FIG{\includegraphics[width=0.9\textwidth]{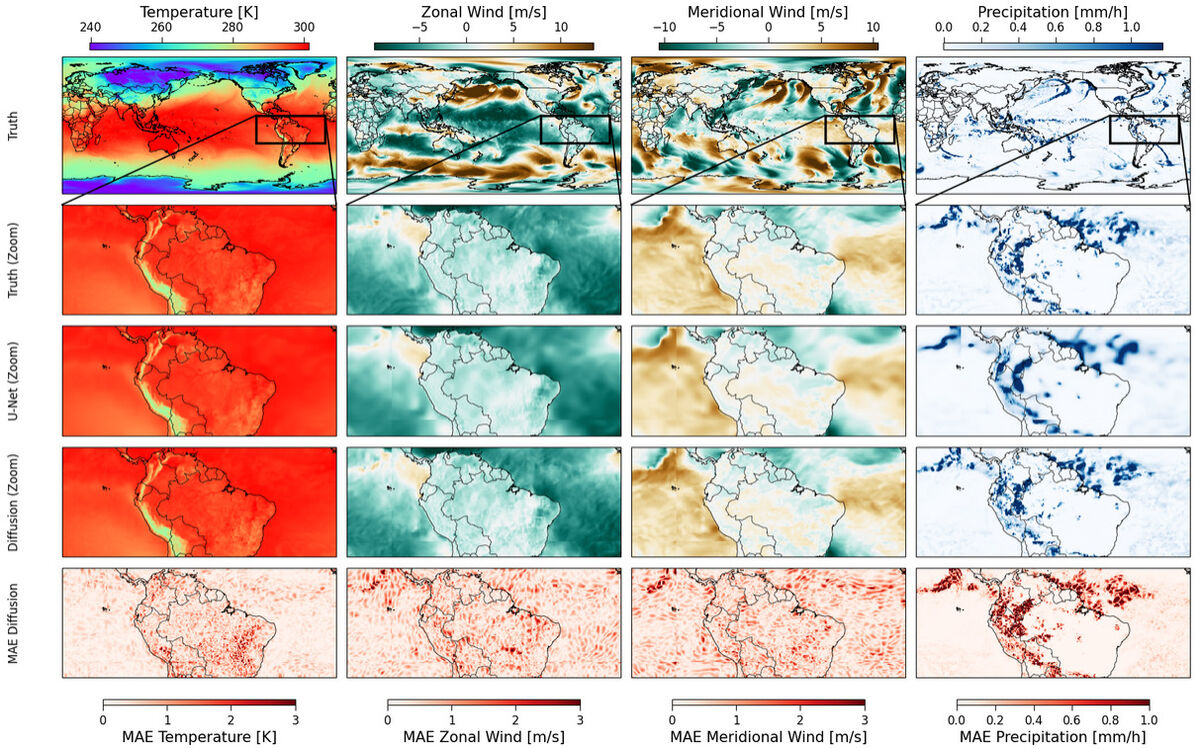}}
{\caption{Surface plots are shown for t2m, u10, v10, and tp (columns 1 to 4). The first row shows the full-domain ground truth. The second to fifth rows present a zoomed-in view over South America, showing respectively the truth, the U-Net prediction, the diffusion model prediction, and the difference between the diffusion model prediction and the truth. All panels correspond to 2021-01-01 06:00 UTC}
\label{zoomin}}
\end{figure}

\subsection{Residual fields}\label{residual_fields}
Figure~\ref{residuals} shows the residual fields (the difference between the HR prediction and the CU input), representing the fine-scale structures restored by the models. The $R^2$ scores of the residuals are apparently lower than those of the full fields (Figure~\ref{scatter_pdf_psd}) due to reduced variance, not degraded performance. The errors (MAE, RMSE) remain unchanged, indicating consistent accuracy, but lower variance of residual results in lower $R^2$ scores. The middle row shows the PDFs of the residuals. The diffusion model generates residuals centered near zero with a dispersion corresponding to the truth variance, while the U-Net produces slightly narrower and sharper PDFs, underestimating fine-scale variability (more visible for the u10, v10, and precipitation). For precipitation, the contrast is more pronounced: the diffusion model produces broader tails, reflecting extreme events, while the U-Net underestimates extreme events. The PSDs (bottom row) show that the diffusion model injects energy at high wavenumbers, producing fine-scale structures, while the U-Net exhibits earlier attenuation at high wavenumbers, confirming a limited capacity to reconstruct the finest scales.

\begin{figure}[t]%
\FIG{\includegraphics[width=0.95\textwidth]{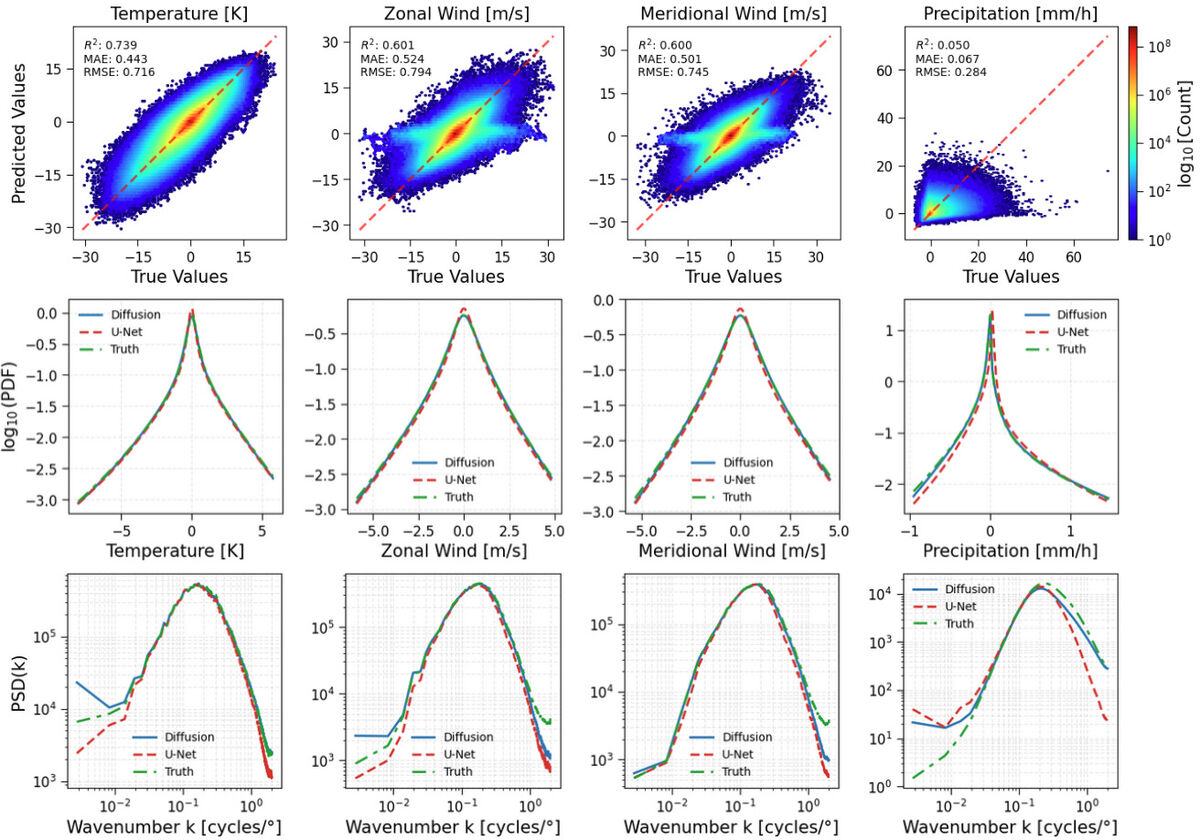}}
{\caption{Evaluation of residual fields for the complete 2021 evaluation dataset. Columns represent t2m, u10, v10, and tp, respectively. Rows display density scatter plots (top), PDFs (middle), and PSDs (bottom)}
\label{residuals}}
\end{figure}

\subsection{Spatial distribution of relative errors}\label{relative_errors}
Figure~\ref{r_error} shows the spatial distribution of errors for t2m, u10, v10, and precipitation. For t2m, the relative error is amplified at high latitudes, particularly over the Antarctic region, and in areas with complex topography such as the Himalayas and the Andes. The (u10, v10) components also show increased relative errors in coastal regions and along strong topographic gradients, reflecting the difficulty of representing near-surface winds influenced by land--sea contrasts. In contrast, precipitation is presented using absolute error, since relative error becomes unreliable for very small values. The largest precipitation errors are concentrated in the tropics, especially along the intertropical convergence zone and over tropical rainforest regions. 

\begin{figure}[htb!]%
\FIG{\includegraphics[width=\textwidth]{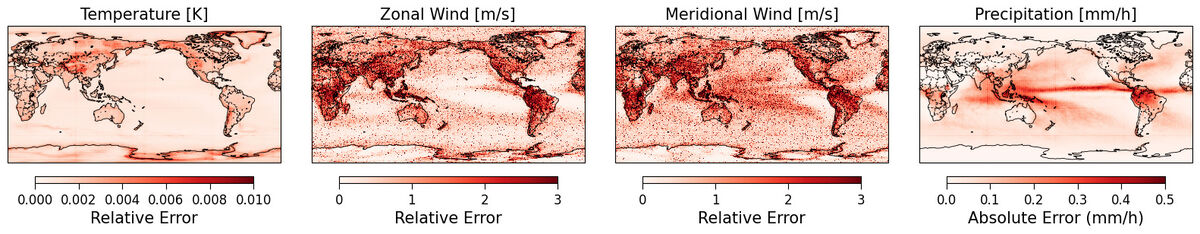}}
{\caption{Spatial distribution of relative error for t2m, u10, and v10 (first three panels), and absolute error for precipitation (rightmost panel)}
\label{r_error}}
\end{figure}

\subsection{Inter-variable correlation analysis}\label{correlation_analysis}
To assess whether the diffusion model preserves physical consistency, we evaluate the temporal co-variability between variable pairs using the Pearson correlation coefficient \citep{vrac_2015}. Capturing realistic multivariate dependencies is particularly important because many environmental and climate impacts depend on compound conditions, such as hot--dry days or wind--precipitation interactions. These relationships strongly influence impact models related to hydrology, agriculture, and wildfire risk \citep{coppola_2021}. For each pair, correlations are computed over time at each grid cell, yielding a spatial map of local coupling strength. Given two variables $y_1(b,h,w)$ and $y_2(b,h,w)$ defined over time $b=1,\dots,B$ and spatial dimensions $h=1,\dots,H$, $w=1,\dots,W$, the temporal Pearson correlation coefficient at each grid cell $(h,w)$ read:

\begin{equation}
{\text{corr}}(h,w) =
\frac{
\sum_{\mathrm b=1}^{\mathrm B} \Big(y_1(b,h,w) - \bar{y}_1(h,w)\Big)\Big(y_2(b,h,w) - \bar{y}_2(h,w))
}{
\sqrt{\sum_{\mathrm b=1}^{\mathrm B} \Big(y_1(b,h,w) - \bar{y}_1(h,w)\Big)^2} \;
\sqrt{\sum_{\mathrm b=1}^{\mathrm B} \Big(y_2(b,h,w) - \bar{y}_2(h,w)\Big)^2}}
\label{eq:temporal_corr}
\end{equation}
where $\bar{y}$ is the temporal mean of a variable. For each pair of variables, three correlation maps are produced: the truth, the diffusion model prediction, and the CU input. 

Figure~\ref{fig:temporal_corr} shows spatial maps of temporal correlations between pairs of variables, with rows corresponding to each variable pair and columns to the truth, prediction, and CU input. The predicted correlations closely match the truth, capturing large-scale positive and negative structures and preserving dominant physical couplings, such as coherent temperature--wind patterns over oceans and continents. The coarse inputs appear smoothed, whereas the diffusion model restores much of this fine-scale structure, indicating a successful reconstruction of local inter-variable relationships. Agreement is strongest for temperature--wind pairs, while precipitation correlations are more heterogeneous, yet still retain key patterns that are absent in the coarse inputs. 

\begin{figure}[t]
    \centering
    \FIG{\includegraphics[width=\textwidth]{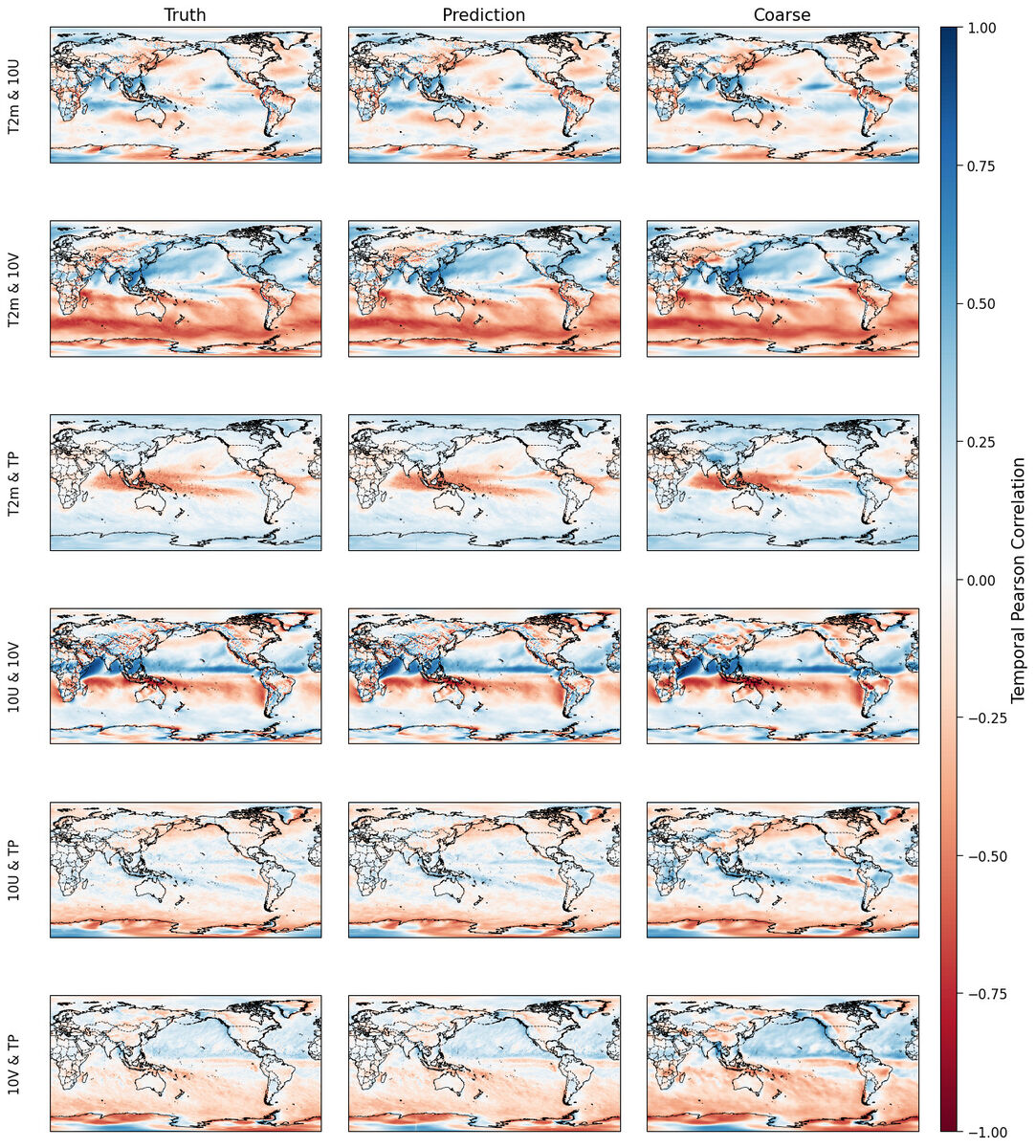}}
    {\caption{Spatial map of the temporal Pearson correlation coefficient between pair variables}
    \label{fig:temporal_corr}}
\end{figure}
}
\end{appendix}

\end{Backmatter}

\end{document}